# Study of multi-neutron systems with SAMURAI spectrometer

Z. H. Yang[1,2,*], F. M. Marqués[3], N. L. Achouri[3], D. S. Ahn[2], T. Aumann[4,5], H. Baba[2],
D. Beaumel[6], M. Böhmer[7], K. Boretzky[5,2], M. Caamaño[8], S. Chen[9], N. Chiga[2],
M.L.Cortés[2], D. Cortina[8], P. Doornenbal[2], C. A. Douma[10], F. Dufter[7], J. Feng[11,2],
B.Fernández-Domínguez[8], Z. Elekes[12,2], U. Forsberg[13,24], T. Fujino[14], N. Fukuda[2],
I.Gašparić[15,2], Z. Ge[2], R. Gernhäuser[7], J. M. Gheller[16], J. Gibelin[3], A. Gillibert[16],
B.M.Godoy[3], Z. Halász[12], T. Harada[17,2], M. N. Harakeh[10], A. Hirayama[18,2],
S.W.Huang[11,2], N. Inabe[2], T. Isobe[2], J. Kahlbow[4,2], N. Kalantar-Nayestanaki[10],
D.Kim[19], S. Kim[19], M. A. Knösel[4], T. Kobayashi[20], Y. Kondo[18], P. Koseoglou[4,5],
Y.Kubota[2], I. Kuti[12], C. Lehr[4,2], P. J. Li[9], Y. Liu[11,2], Y. Maeda[21], S. Masuoka[22],
M.Matsumoto[18,2], J. Mayer[23], H. Miki[18], M. Miwa[17,2], I. Murray[2], T. Nakamura[18],
A.Obertelli[4], N. Orr[3], H. Otsu[2], V. Panin[2], S. Park[19], M. Parlog[3], S. Paschalis[4,13],
M.Potlog[25], S. Reichert[7], A. Revel[26], D. Rossi[4], A. Saito[18], M. Sasano[2], H. Sato[2],
H.Scheit[4], F. Schindler[7], T. Shimada[18], Y. Shimizu[2], S. Shimoura[22], I. Stefan[6],
S.Storck[4], L. Stuhl[22], H. Suzuki[2], D. Symochko[4], H. Takeda[2], S. Takeuchi[18],
J.Tanaka[4,5], Y. Togano[14,2], T. Tomai[18,2], H. T. Törnqvsit[4,2]. J. Tscheuschner[4],
T.Uesaka[2], V. Wagner[4], K. Wimmer[22], H. Yamada[18], B. Yang[11,2], L. Yang[22],
Y.Yasuda[18,2], K.Yoneda[2], L. Zanetti[4,2], J. Zenihiro[2]

[1] Research Center for Nuclear Physics, Osaka University, Ibaraki, Osaka 567-0047, Japan
[2] RIKEN Nishina Center, 2-1 Hirosawa, Wako, Saitama 351-0198, Japan
[3] LPC Caen, ENSICAEN, Université de Caen, CNRS/IN2P3, F-14050 CAEN Cedex, France
[4] Institut für Kernphysik, Technische Universität Darmstadt, D-64289 Darmstadt, German
[5] GSI Helmholtzzentrum für Schwerionenforschung, 64291 Darmstadt, Germany
[6] Institut de Physique Nucléaire Orsay, IN2P3-CNRS, 91406 Orsay Cedex, France
[7] Technische Universität München, 85748 Garching, Germany
[8] Departamento de Física de Partículas and IGFAE, Universidade de Santiago de Compostela, E-15782 Santiago de Compostela, Spain
[9] Department of Physics, The University of Hong Kong, Pokfulam Road, Hong Kong, China
[10] KVI-CART, University of Groningen, Zernikelaan 25, 9747 AA Groningen, The Netherlands
[11] School of Physics and State Key Laboratory of Nuclear Physics and Technology, Peking University, Beijing 100871, China
[12] MTA ATOMKI, 4001 Debrecen, Hungary
[13] Department of Physics, University of York, York YO10 5DD, United Kingdom
[14] Department of Physics, Rikkyo University, Toshima, Tokyo 172-8501, Japan
[15] Ruđer Bošković Institut (RBI), Zagreb, Croatia
[16] CEA, Centre de Saclay, IRFU, F-91191 Gif-sur-Yvette, France
[17] Toho University, Tokyo 143-8540, Japan
[18] Department of Physics, Tokyo Institute of Technology, Meguro, Tokyo 152-8551, Japan
[19] Department of Physics, Ehwa Womans University, Seoul, Korea
[20] Department of Physics, Tohoku University, Miyagi 980-8578, Japan
[21] Department of Applied Physics, University of Miyazaki, Miyazaki 889-2192, Japan
[22] Center for Nuclear Study, University of Tokyo, 2-1 Hirosawa, Wako, Saitama 351-0198, Japan

[23] Institut für Kernphysik, Universität zu Köln, Köln, Germany
[24] Department of Physics, Lund University, 22100 Lund, Sweden
[25] Institute of Space Sciences, Magurele, Romania
[26] Grand Accélérateur National d'Ions Lourds (GANIL), CEA/DRF-CNRS/IN2P3, Bvd Henri Becquerel, 14076 Caen, France
*zhyang@ribf.riken.jp

**Abstract.** The tetraneutron has been drawing the attention of the nuclear physics community for decades, but a firm conclusion on its existence and properties is still far from being reached despite many experimental and theoretical efforts. New measurements have recently been performed at RIBF with the SAMURAI spectrometer by applying complementary reaction probes, which will help to pin down the properties of this four-neutron system.

# 1   Introduction

There has been a long-standing question among the nuclear physics community whether a nucleus made purely of neutrons without any protons can exist or not. These so-called "*Neutral nuclei*" have attracted in particular a lot of attention over the past decades. These multi-neutron systems, whether existing as bound or quasi-bound (resonant) states, have fundamental importance in nuclear physics. They provide the possibility to investigate "purely" the nuclear forces free from Coulomb interaction, which is essential for developing the nuclear theory, and critical for our understanding of neutron-rich nuclear matter and neutron stars. The two-neutron system, dineutron ($^2n$), has been well known to be unbound. For the tetraneutron ($^4n$), however, no firm conclusion has been drawn yet despite many experimental and theoretical efforts.

Earlier experimental attempts to search for a bound $^4n$ with a wide variety of methods all failed to find positive evidence. The existence of a bound state of four neutrons has been ruled out by calculations based on standard nuclear forces. But the possibility for $^4n$ existing as a resonant state is supported by some theoretical calculations. In 2002 Marqués *et al.* reported the possible existence of a bound or low-lying resonant $^4n$ state [1]. The resurgence of interest on this topic in recent years was triggered by the report on the observation of a low-lying $^4n$ resonance (although the resolution made the result also compatible with a bound state within error bars) released by Kisamori *et al.* in 2016. This concerned a background-clean measurement with the SHARAQ spectrometer, but only four events were identified indicative of a "candidate" $^4n$ resonant state [2]. This observation was qualitatively reproduced by recent Quantum Monte Carlo (QMC) [3] and No-Core Shell Model calculations [4], but was not supported by some other *ab-initio-type* calculations [5,6].

New experimental measurements on $^4n$ have been recently performed at RIKEN Radioactive Isotope Beam Factory (RIBF) facility: updated double-charge exchange reaction $^4\text{He}(^8\text{He},\alpha\alpha)^4n$ by Shimoura *et al.* (for details see report by Shimoura in the present proceedings), $^8$He$(p,p\alpha)^4n$ with inverse kinematics by Rossi and Paschalis *et al.*, and $^8$He$(p,2p)^7$H$\{t+^4n\}$ with inverse kinematics by Yang and Marqués *et al.*. These experiments will provide independent and complementary information about $^4n$, considering the different population processes, resolutions, production yields and analysis methods, which will help to pin down the properties of this four-neutron system. The latter two experiments, both requiring high luminosity from the thick liquid hydrogen target MINOS [7] and the high neutron detection efficiency of the combined neutron detection array NeuLAND+NEBULA, were arranged in a campaign with the SAMURAI spectrometer in 2017. In the present report we will focus on the $^8$He$(p,2p)$ experiment, and a brief introduction of the $^8$He$(p,p\alpha)$ experiment will also be presented.

## 2   Description of the  $^8$He$(p,2p)^7$H$\{t+4n\}$ experiment

The $^8$He$(p,2p)^7$H$\{t+4n\}$ experiment was carried out at the RIBF, which is operated by the RIKEN Nishina Center and the Center for Nuclear Study (CNS), University of Tokyo. The secondary $^8$He beam with an energy of ~ 150 MeV/nucleon and an intensity of ~$10^5$ pps was produced in the fragmentation of $^{18}$O on the $^9$Be primary target, and then purified and transported by the BigRIPS beam line. The incident $^8$He particles were identified event-by-event with TOF-ΔE method by using plastic scintillators on the beam line and tracked with two multi-wire drift chambers (BDC1 and BDC2) onto the 150-mm-thick vertex-tracking liquid hydrogen target MINOS. $^7$H was then produced from the $(p,2p)$ reaction off $^8$He. A schematic view of the experimental setup is presented in Fig. 1.

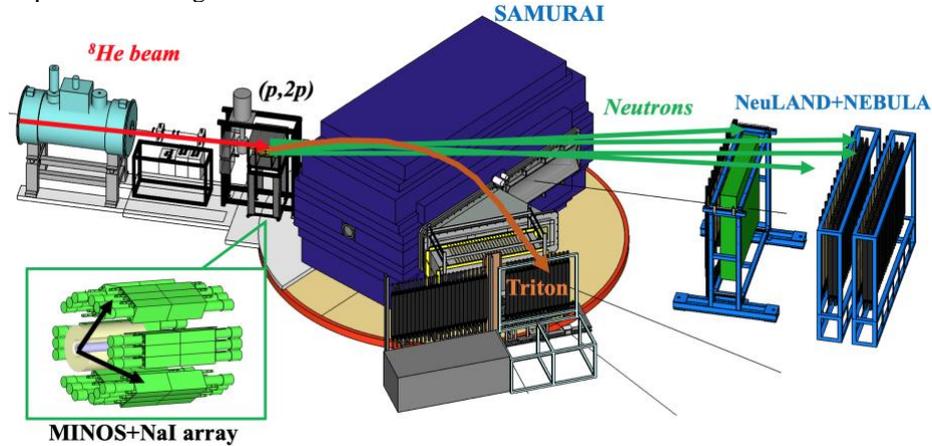

Fig. 1 Schematic view of the setup for the $^8$He$(p,2p)^7$H$\{t+4n\}$ experiment with inverse kinematics.

The recoil protons from the $(p,2p)$ reaction were tracked by the TPC of MINOS and then recorded by a compact NaI array surrounding the target, constructed with 36 crystals from the DALI2 in-beam gamma-ray spectrometer [8] and arranged into two symmetric rings. The energy resolution of the NaI scintillators was determined to be around 1% (FWHM) in measurements with 80-MeV protons at CYRIC of Tohoku University, and an energy calibration was performed by measuring the proton-proton elastic scattering at 175 MeV with the same setup. The charged fragments were analyzed by the SAMURAI spectrometer [9]. Coincident detection of the multiple decay neutrons is critical but extremely challenging for studies of these multi-neutron systems. The NeuLAND demonstrator with four double-planes from GSI [10] was added to the existing NEBULA array, which provides the highest 4-neutron detection efficiency ($\varepsilon_{4n}$ ~1%) achievable at present. It is worthwhile to note that all the reaction products were recorded in the present experiment, providing the complete kinematics of the reaction and therefore eliminating possible ambiguity from identification of multi-neutron events. Another advantage of the present kinematically complete measurement is the applicability of the so-called "*Missing-Invariant-Mass method*", which requires the detection of only three of the four neutrons by reconstructing the kinematics of the missing neutron from the remaining particles and therefore largely enhances the sensitivity in the vicinity of the threshold (by a factor of ~20).

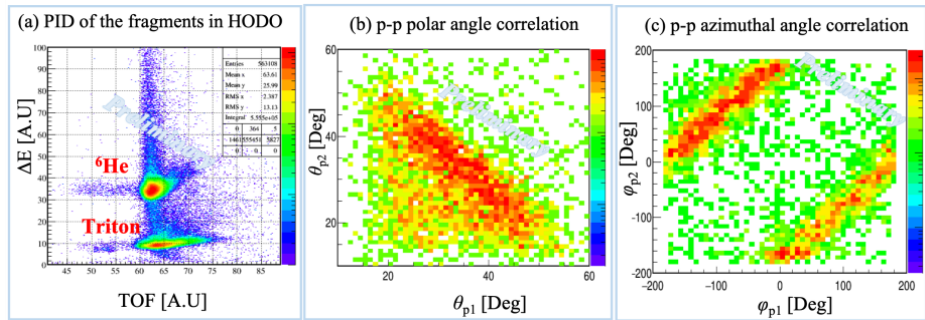

Fig.2. (a) PID of fragments by measuring ToF and ΔE with HODO. Polar angle (b) and azimuthal angle (c) correlations of the two recoil protons populated in the $^8$He(p,2p) reaction.

The charged fragments are identified from the TOF and ΔE signals from the HODO plastic scintillator array at the exit of SAMURAI. As shown in Fig.2(a), tritons are well recorded and clearly separated from $^6$He fragments. Angular information of the recoil protons is provided by the TPC surrounding the target. In Fig.2(b) and (c), the correlations of polar angles and azimuthal angles of the reoil proton pair in coincidence with outgoing triton fragments are presented, respectively. An opening angle of ~70 degrees and the coplanarity are evidently observed, in agreement with the expected correlation pattern for the quasi-free (*p*,2*p*) reaction off $^8$He. We have also checked the registered neutron multiplicity in the NeuLAND+NEBULA array by applying causality cuts on the space-time separation of neutron hits in order to remove fake multi-neutron events (due to the so called cross-talk). In this way, we estimate that the recorded full-kinematics events will be of the order of 50k, which will permit a detailed investigation of the tetraneutron system.

## 3   Perspectives

Within the same campaign, $^8$He($p, p\alpha$)4$n$ was measured by Rossi and Paschalis *et al.*, taking advantage of the well developed $\alpha+4n$ cluster structure of $^8$He. The final-state interaction with the charged fragments is minimized by carrying out the measurement at very backward scattering angles in the center of mass. The experimental setup is similar to the $^8$He($p,2p$)$^7$H$\{t+4n\}$ experiment (Fig. 1). Instead of the TPC+NaI detectors, a silicon-tracker system was introduced at the target region for the tracking of protons and alpha particles [11]. The data analysis is still in progress, but protons and alpha particles are clearly identified in coincidence, and the four-neutron system can then be reconstructed from the missing-mass method.

In 2018, the first direct experimental investigation of the 6-neutron system ("hexaneutron") was made by Beaumel *et al.* by measuring the $^{14}$Be($p,p\alpha\alpha$) reaction with SAMURAI spectrometer. Furthermore, some new proposals on multi-neutron systems based on ($p,2p$) and ($p,p\alpha$) reactions are also in preparation.

There are in general two key yet challenging factors for experimental studies of these exotic multi-neutron systems. One is the population process, and the other is the detection and identification of the multi-neutron events. It is not clear at present which reaction will selectively populate the multi-neutron systems we are interested in, and complementary measurements with different reaction probes would in this sense help to reach a definite conclusion. The high-quality secondary beams provided by RIBF together with the large-acceptance SAMURAI spectrometer and the associated large neutron detector arrays (including the EXPAND upgrade project for NEBULA) are providing new opportunities for sophisticated studies of these multi-neutron systems.

We acknowledge the support of the RIBF accelerator staff and the BigRIPS team for providing the high-quality beam. Z. H. Yang acknowledges the financial support from the Foreign Postdoctoral Researcher program of RIKEN. I. G. was supported by HIC for FAIR and Croatian Science Foundation Project No. 7194.